# Automatic Hardware Synthesis for a Hybrid Reconfigurable CPU Featuring Philips CPLDs

Bernardo Kastrup

*Philips Research Laboratories, Prof. Holstlaan 4 (WL11), 5656 AA Eindhoven,
The Netherlands. Tel.: +31 40 274 4421. Fax: +31 40 274 4004.
E-mail: kastrup@natlab.research.philips.com*

**Abstract**. A high-level architecture of a hybrid reconfigurable CPU, based on a Philips-supported core processor, is introduced. It features the Philips XPLA2 CPLD as a reconfigurable functional unit. A compilation chain is presented, in which automatic implementation of time-critical program segments in custom hardware is performed. The entire process is transparent from the programmer's point of view. The hardware synthesis module of the chain, which translates segments of assembly code into a hardware netlist, is discussed in details. Application examples are also presented.

## 1 Introduction

The integration of reprogrammable logic and a core CPU in a single die indicates a new computing paradigm which ultimately aims at joining the flexibility of a general-purpose processor and ASIC-comparable performance [1, 2]. The reprogrammable logic is used to speed-up time-critical segments of the application, while the core CPU remains responsible for the main program sequencing and control. A couple of factors (amongst others) must be considered when thinking of such a hybrid reconfigurable processor:

- The granularity of the reprogrammable logic;
- How coupled is its integration with the CPU.

Fine-grained reprogrammable logic is well suited for sparse bitwise computations. It has poor performance, however, for regular wordwise computations. FPGA-based full-word multipliers, for instance, may be up to two orders of magnitude slower than their equivalent dedicated logic implementations [3]. On the other hand, coarse-grained reprogrammable logic is less flexible and cannot benefit from bit-level parallelism.

While the granularity of the reprogrammable logic itself determines how well it may perform for different kinds of computation, the level of its integration with the core CPU determines the granularity of the application segments which it can execute. For instance, a coprocessor-like approach is decoupled [4, 5]. Core CPU and reconfigurable coprocessor can work in parallel, which is suitable for accelerating time-demanding program functions operating asynchronously with the rest of the computation. Ideally, the application segments implemented in the coprocessor compute large chunks of data in memory, writing results directly back to memory.

On the other hand, when the reprogrammable logic is integrated within the core CPU's data-path, as illustrated in Figure 1 for a typical RISC pipeline, small segments of the application, highly coupled with the rest of the computation, may be implemented in custom hardware as an application-specific instruction.

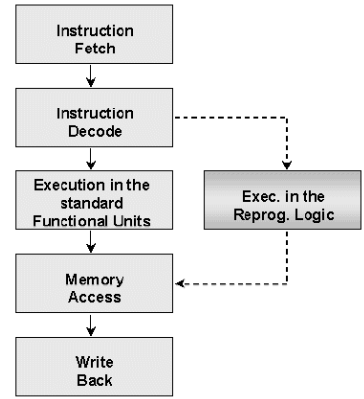

**Figure 1. Datapath integration between reprogrammable logic and core CPU, for a typical RISC pipeline.**

In this paper, we present a high-level architecture definition of a hybrid reconfigurable processor based on a Philips-supported CPU core. A data-path integration approach was chosen, and the Philips XPLA2 CPLD architecture was the fine-grained reprogrammable hardware platform used. Applications that are biased towards sparse and highly coupled bitwise computation (*e.g.* graphics) are the target.

Section 2 describes the XPLA2 CPLD architecture. Section 3 introduces the Philips Hardware Description Language (PHDL), and related software tools, used to program Philips CPLDs. Section 4 discusses the hybrid architecture proposed, that merges a standard processor core with XPLA2 CPLD-based functional units. A compilation strategy targeting the hybrid architecture is introduced in Section 5, and its hardware synthesis part is presented in Section 6. Section 7 illustrates some application examples of the hardware synthesis module, and Section 8 summarizes and concludes this work.

## 2 Philips CoolRunner CPLDs

The new generation of Philips CoolRunner CPLDs[1] is based on the XPLA2 architecture[2], which is constructed from so-called Fast Modules that are connected together by a Global Zero Power Interconnect Array (or GZIA). Within each Fast Module there are four Logic Blocks of 20 macrocells each. Each Logic Block contains a PAL structure with four dedicated product terms for each macrocell. In addition, each Logic Block has 32 product terms in a PLA structure that can be shared through a fully programmable OR array to any of the 20 macrocells [6]. Figure 2 illustrates the XPLA2 Logic Block architecture.

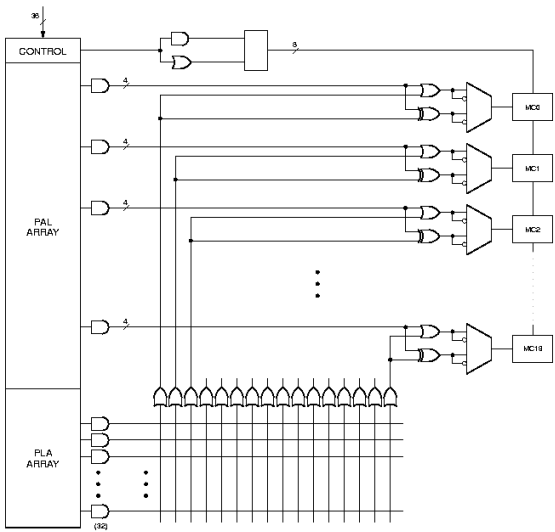

**Figure 2. XPLA2 Logic Block architecture (extracted from [6]).**

By replacing conventional sense amplifiers methods for implementing product terms with a cascaded chain of pure CMOS gates[3], both stand-by and dynamic power are sensibly reduced in the CoolRunner CPLDs, without sacrificing performance.

---

1. http://www.coolpld.com/
2. Patent pending.
3. The patented *Fast Zero Power* (FZP) technology.

The PZ3960 device, member of the XPLA2 family, offers pin-to-pin propagation delays of 7.5ns through the PAL array. If the PLA array is used, an additional 1.5ns is added to the delay. The GZIA adds a second fixed propagation delay of 4.0ns. The PZ3960 is SRAM-based, configured at power-up, and offers 384 I/O pins.

## 3 PHDL at a Glance

The XPLA Designer tools used to program the CoolRunner CPLDs support design definition, functional simulation, device fitting and post layout (timing) simulation. It accepts design entries written in PHDL (Philips Hardware Description Language).

An example PHDL design is shown in Figure 3. A standard header contains the module name and the title of the design. A declarations section declares input and output pins (the last ones specified as 'com' pins, or 'combinatorial'), as well as sets, collections of pins which can be referenced all together. For instance, the line 'R24 = [R24b31..R24b0]' declares the set R24 as including the pins from R24b31 to R24b0 (32 pins).

```
module  seg1
title   'seg1.phd'

declarations
R24b31..R24b0       pin;
R24 = [R24b31..R24b0];
Rout24b31..Rout24b0 pin istype 'com';
Rout24 = [Rout24b31..Rout24b0];
TEMP1b31..TEMP1b0   node istype 'com';
TEMP1 = [TEMP1b31..TEMP1b0];
...
TEMP8b31..TEMP8b0   node istype 'com';
TEMP8 = [TEMP8b31..TEMP8b0];

equations
TEMP1 = [R24b7, ..., R24b0, 0, ..., 0];
TEMP2 = R24 & 65280;
TEMP3 = [TEMP2b23, TEMP2b22,..., TEMP2b1,
        TEMP2b0, 0, ..., 0];
TEMP4 = TEMP1 + TEMP3;
TEMP5 = [0, 0, 0, 0, 0, 0, 0, 0, R24b31,
        R24b30, ..., R24b8];
TEMP6 = TEMP5 & 65280;
TEMP7 = TEMP4 + TEMP6;
TEMP8 = [0, ..., 0, R24b31, R24b30, ...,
        R24b24];
Rout24 = TEMP7 + TEMP8;
end
```

**Figure 3. PHDL design example (ellipses '...' are added for compactness).**

Intermediate nodes ('node'), which hold temporary results, must also be declared, but may (should) be collapsed during design compilation for optimization purposes. In the 'equations' section, the actual design behaviour is specified by means of high-level (and readable) statements.

The PHDL compiler optimizes the circuit specification in terms of performance and resource usage. Precise timing and fitting reports are also produced by the XPLA Designer tools in the form of text files.

## 4 A Hybrid RISC Architecture

The integration of reprogrammable hardware within the datapath of a core CPU is no news in the configurable computing world [7-9]. What makes our approach different, though, is that, instead of designing a new architecture for the reprogrammable hardware or the core CPU, we aim at putting together existing commercial architectures.

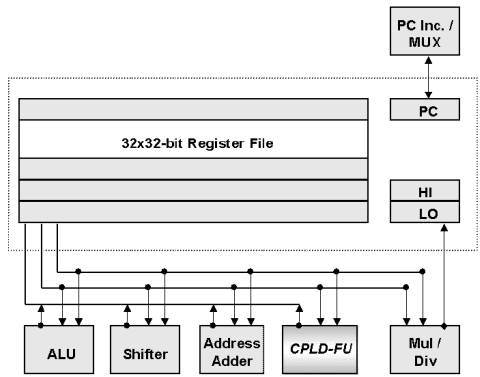

**Figure 4. The hybrid architecture.**

Philips currently supports 2 lines of RISC processors: Philips RISC and TriMedia[1]. In this work, we will base our high-level architecture on the Philips RISC, a MIPS processor [10]. Figure 4 illustrates the hybrid architecture proposed. A PZ3960 CPLD-based functional unit (CPLD-FU) is added in parallel to the standard MIPS-2 FUs. More than one CPLD-FU can be added. Even though such an embedded application would require minor adaptations (and down stripping) on the CPLD architecture, this will not be addressed in this work.

The custom CPLD-FU instruction is a three-operand register-type operation added to the instruction set:

```
cpld   rd, rs, rt
```

---
1. http://www.semiconductors.philips.com/trimedia/

Where `rd` is the destination register.

## 5 Compiling Applications

A compilation chain to target our hybrid CPU architecture is proposed in Figure 5. It performs the following steps:

1. Source code (typically written in C) is processed by an existing core compiler, generating assembly code;
2. The assembly code is parsed by a candidates detection/selection module which looks for instruction sequences, within basic blocks [11], suitable for implementation in the reprogrammable logic resources. At this point, profile information is used to identify the time-critical segments of the application;

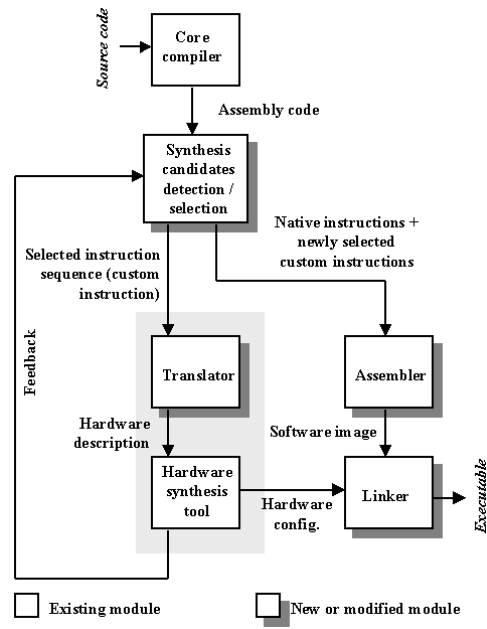

**Figure 5. Compilation chain for the hybrid CPU.**

3. Each instruction sequence selected is then translated, via an automatic translator, into a hardware description in PHDL;
4. An existing hardware synthesis tool, the XPLA Designer, reads the hardware description, compiles, optimizes and fits it into the CPLD-FU architecture. Fitting and timing analysis reports are generated;
5. Timing and fitting information is then fed back to the candidates detection/selection module. If the implementation has a too long delay or doesn't physically fit into the CPDL-FU, a new instruction sequence is selected and the cycle restarts;

6. The finally selected instruction sequence(s) are replaced, in the assembly code, with the newly synthesized CPLD-FU instructions. The whole set of instructions (native + custom instructions) is then sent to a modified assembler;

7. The hardware image (the netlist generated by the hardware synthesis tool) of the application is combined by a linker with the software image generated by the assembler, producing an executable.

## 6 From Assembly to Hardware

Currently, work on the compilation chain proposed in Figure 5 is still in progress. The gray frame, nevertheless, has been already completed.

The module 'translator' is a program which reads a text file containing a selected assembly segment and translates it into a PHDL-equivalent. Figure 6 illustrates the simplified software architecture of the module. The assembly segment must be a straight line of code (basic block [11]) with one or two inputs (operands) and one output (result). The current supported opcodes for MIPS assembly are: SUBU, ADDU, AND, OR, SRL, SLL, SRA and LI. This set covers the basic arithmetic and logic functions which CPLD implementations are not excessively costly in terms of latency and/or hardware resources usage. Variable length shifts, as well as multiplies and divides, on the other hand, are fairly costly, and have not been implemented.

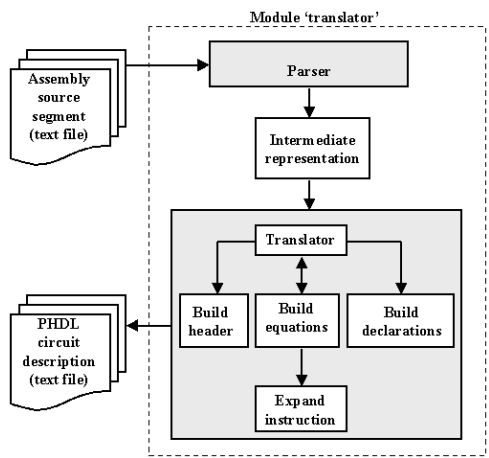

**Figure 6. The simplified architecture of the module 'translator'.**

The assembly segment is initially read and translated into a linked list-based intermediate representation by a parser. Immediate values are differentiated from register-based operands. An assembly instruction writing a given general-purpose register is linked to the instruction(s) using that register as operand. The instruction opcodes are replaced by internal operations (specified in a list of supported operations), such that the intermediate representation abstracts the specific assembly implementation it is derived from.

After the intermediate representation is built, the actual translation may start. The header of the PHDL file is first built and written. It contains the module name and the title of the design. The control is then transferred to the equations builder. For each internal operation in the intermediate representation, the equations builder determines its operands (input pins or intermediate nodes - see Section 3 - in the PHDL file) and where its result should be written to (output pins or, still, intermediate nodes). An instruction expander then receives the operands/result information and generates PHDL code corresponding to the internal operation. After all equations are generated, the number of intermediate nodes required is returned. This number is used in a declarations builder, which writes in the PHDL file all of the input/output pins declarations necessary, as well as the nodes declarations.

The hardware synthesis tool is the XPLA Designer. A shell script coordinates the interfacing between the translator and the hardware synthesis tool, such that a single command brings the MIPS assembly segment directly to its hardware netlist for the XPLA2 architecture, as well as produces fitting and timing reports.

## 7 Application Examples

Some examples of the translator use for a couple of applications will be shown. We have run the 'triangles' software, a graphics application internally used at Philips for benchmarking purposes, and the arithmetic module of the LIFE Linker, by Ross Morley.

1. The compilation of 'triangles' for MIPS generates the following segment of code:

```
and     $8, $9, 1
li      $10, 1
subu    $11, $10, $8
sll     $12, $11, 1
```

Its automatic translation to PHDL gives:

```
declarations
R9b31..R9b0 pin;
R9 = [R9b31..R9b0];
Rout12b31..Rout12b0 pin istype 'com';
Rout12 = [Rout12b31..Rout12b0];
TEMP1b31..TEMP1b0 node istype 'com';
TEMP1 = [TEMP1b31..TEMP1b0];
TEMP2b31..TEMP2b0 node istype 'com';
TEMP2 = [TEMP2b31..TEMP2b0];
```

```
equations
TEMP1 = R9 & 1;
TEMP2 = 1 - TEMP1;
Rout12=[TEMP2b30, ...,TEMP2b0, 0];
end
```

Where the header was omitted. Note that the li instruction is bypassed, and its immediate is directly forwarded to the subtraction. The netlist generated by compiling the PHDL description leads to the circuit shown in Figure 7.

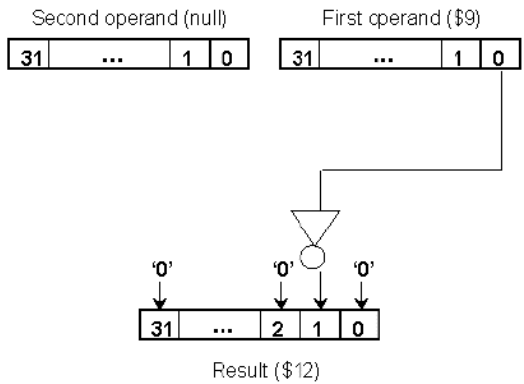

**Figure 7. Example generated CPLD-FU configuration.**

The latency[1] of the implementation is 11.5ns in a Philips PZ3960 CPLD. Therefore, the four assembly instructions can be replaced by:

```
cpld    $12, $9, $0
```

which can execute in one clock cycle, provided that the clock frequency is no greater than 85MHz (ignoring clock overheads). Power consumption savings are also evident, specially with the FZP$^{TM}$ technology of the CoolRunner CPLDs, however not estimated here. Another "side-effect" benefit is a reduction in the register pressure because of the elimination of some temporaries.

2. Another segment extracted from 'triangles' follows:

```
sll $15,$24,24
and $14,$24,0xff00
sll $14,$14,8
addu $15,$15,$14
srl $14,$24,8
and $14,$14,0xff00
addu $15,$15,$14
```

---

1. Throughout this work, the "latency" of a circuit implemented in a Philips CoolRunner CPLD refers to the maximum delay from input to output pin ($Tpd_{max}$), as measured with the timing analyser of the XPLA Designer V2.70.

```
srl $24,$24,24
addu $24,$15,$24
```

This set of nine MIPS assembly instructions merely performs an endian conversion. Its automatically generated PHDL translation is shown in Figure 3. The compiled/optimized equations do not use a single logic operation, just rewiring input to output pins. The latency of the CPLD implementation is also 11.5ns (PAL + GZIA delay). This clearly shows that the PHDL compiler is able to optimize high-level hardware descriptions, automatically generated from assembly code, into optimal circuit configurations.

3. The last example is extracted from the arithmetic module of the LIFE Linker:

```
addu    $14, $5, -1
and     $15, $14, 255
sra     $24, $15, 3
addu    $25, $24, 1
```

Its PHDL translation, as generated by the module 'translator', follows:

```
declarations
R5b31..R5b0 pin;
R5 = [R5b31..R5b0];
Rout25b31..Rout25b0 pin istype 'com';
Rout25 = [Rout25b31..Rout25b0];
TEMP1b31..TEMP1b0 node istype 'com';
TEMP1 = [TEMP1b31..TEMP1b0];
TEMP2b31..TEMP2b0 node istype 'com';
TEMP2 = [TEMP2b31..TEMP2b0];
TEMP3b31..TEMP3b0 node istype 'com';
TEMP3 = [TEMP3b31..TEMP3b0];

equations
TEMP1 = R5 + (-1);
TEMP2 = TEMP1 & 255;
TEMP3 = [TEMP2b31, TEMP2b31, TEMP2b31,
TEMP2b31,  TEMP2b30,  TEMP2b29,   ...,
TEMP2b3];
Rout25 = TEMP3 + 1;
end
```

The header is once more omitted. Because of the subtraction, a denser circuit is generated, which utilizes 38 macrocells and takes 25ns of latency in the PZ3960.

These examples point out some features of our approach:

- The translation from MIPS assembly to PHDL is simple and straight forward. The automatic translator resembles a compiler, with its parsing and conversion phases;

- No optimization at all is done during the translation. The optimizing steps are performed by the XPLA Designer PHDL compiler during hardware compilation. The examples show that a highly optimized circuit is achieved even when very high-level assembly-like circuit descriptions are the input.

## 8 Conclusions and Future Work

We have developed an approach for compiling applications for a hybrid reconfigurable MIPS-based CPU with reprogrammable functional units. The hardware synthesis phase of our compilation chain has been already implemented and is able to compile (segments of) basic blocks of MIPS assembly code, with standard arithmetic and logic operations, into circuit netlists for the Philips XPLA2 CPLD architecture. Unlike other similar approaches, we integrate existing commercial products to achieve our goals, which sensibly reduces development time, risks, and increases the reliability of the system. In this context, Philips XPLA2 CPLDs are used as reprogrammable functional units connected to a standard (also Philips supported) MIPS-2 CPU. For hardware synthesis, a translation from MIPS assembly to PHDL (Philips Hardware Description Language) is performed, and the existing (commercial) PHDL compiler is integrated into the chain to generate the final circuit netlist.

Future steps include the development of the synthesis candidates detection/selection module, such that the entire proposed compilation chain is completed, and a cycle-true simulator for our hybrid CPU. Later on, we plan to extend our studies towards Philips TriMedia, a VLIW media processor, and further investigate the use of reprogrammable hardware in a reconfigurable coprocessor approach.

### Acknowledgements

During the development of this work, fruitful discussions were held with Paul Gorissen, Menno Treffers and Joachim Trescher. Special thanks to Jan Hoogerbrugge, for his constant support and valuable suggestions.